\documentclass[pra,
                twocolumn,
                preprintnumbers,
                amsmath,
                amssymb,
                showpacs,
                superscriptaddress,
                longbibliography]{revtex4-2}
\usepackage{graphicx} 
\usepackage{dcolumn}  
\usepackage{bm}       
\usepackage{amsfonts}
\usepackage[colorlinks=true,linkcolor=blue,citecolor=red,urlcolor=blue]{hyperref}
\usepackage{color}
\usepackage{appendix}
\usepackage{amssymb}
\usepackage[dvipsnames]{xcolor}
\usepackage{soul}

\usepackage{braket}
\usepackage{amsmath}
\usepackage[normalem]{ulem}
\newcommand{\eq}[1]{Eq.\,(\ref{#1})}

\begin{document}

\title{Imaging atomic scattering potential in centroidal diffraction of elastic electrons}

\author{R. Aiswarya}
\affiliation{Department of Physics, Indian Institute of Technology Patna, Bihar 801103, India}

\author{Jobin Jose}
\email[]{jobin.jose@iitp.ac.in}
\affiliation{Department of Physics, Indian Institute of Technology Patna, Bihar 801103, India}

\author{Nenad Simonovi\'{c}}
\affiliation{Institute of Physics Belgrade, Pregrevica 118, Belgrade, Serbia}

\author{Bratislav P. Marinkovi\'{c}}
\email[]{bratislav.marinkovic@ipb.ac.rs}
\affiliation{Institute of Physics Belgrade, Pregrevica 118, Belgrade, Serbia}

\author{Himadri S. Chakraborty}
\email[]{himadri@nwmissouri.edu}
\affiliation{School of Natural Sciences, D.L.\ Hubbard Center for Innovation, Northwest Missouri State University, Maryville, Missouri 64468, USA}

\begin{abstract}
{The Fraunhofer diffraction of quantum particles from materials with sharp electron-density edges or symmetric bond structures is ubiquitous. In contrast, diffraction from atoms with characteristic asymptotically-diffused electron distribution is far less intuitive, although known for many years. The current study unravels an unusual diffraction mechanism of elastic electrons from diffused atomic diffractors. Consequently, the fringe pattern converted to the Fourier reciprocal space maps out the effective scattering potential, which is {\em not} accessible in direct measurements. This may benefit large-volume and multi-scale computational modeling of processes in materials where adopting the complex exact potentials is challenging and expensive. The study employs relativistic partial wave analysis with atoms modeled in the Dirac-Fock formalism and performs \textit{e}-Cd measurements in absolute scale. Analysis for Mg, Ba, and Ra targets demonstrates the universality of the mechanism.} 
\end{abstract}

\maketitle
Varieties of structural information of matter are eminently extractable by diffracting charged particles (electrons or positrons) off materials. For systems more complex than atoms, it is straightforward to understand the bending of projectile plane wave into spherical waves from the well-defined edges of the delocalized electron-cloud and from the localized atomic or lattice centers. This results in the overlap of spherical waves with specific path-differences leading to diffraction features whose Fourier transform (FT) can reveal real-space structural information. For electron projectiles, capturing the scattering signal by a charge-coupled-device detector and the transmission signal by a transmission electron microscope are at the center of Fraunhofer-type diffraction studies. Examples include studies of molecules~\cite{saha2022,amini2020,marinkovic2015development,dampc2007differential}, nanostructures~\cite{ponce2021}, surfaces~\cite{Kienzle2012}, and crystals~\cite{gemmi2019}. Recently, the \textit{e}-C$_{60}$ scattering study~\cite{aiswarya2024simultaneous} has shown that diffraction effects can be captured in two modes, (i) in the scattering angular distribution with a fixed impact energy ($E$) or momentum and (ii) in the impact energy distribution at a fixed scattering direction, with possible benefits in ultrafast electron diffraction research.

The first experimental evidence of diffraction in scattering by simple gas-phase atoms was detected a century ago by Dymond~\cite{dymond1927} and Bullard \& Massey~\cite{bullard1931}. Today, crossed-beam electron scattering has become increasingly multi-dimensional, combining kinematic coincidence detection~\cite{jahnke2021}, ultrafast time-resolution~\cite{chapenois2023}, control over target orientation/state~\cite{lee2024}, and extension to complex systems~\cite{dinger2025}. These advances are crucial for accessing electron correlation, scattering dynamics, and quantum many-body phenomena. Meanwhile, powerful methods, like partial wave analysis~\cite{joachain1975quantum} and quantum defect theory~\cite{seaton2019quantum}, are formulated to describe low- and high-energy collisions and resonances. Later, Dirac partial wave analysis included relativistic effects~\cite{wavemech_applications,mott1933theory}. Techniques, such as, close-coupling theory~\cite{stelbovics1990scattering}, R-matrix~\cite{descouvemont2010r}, relativistic distorted wave approximation~\cite{zuo1991relativistic}, distorted-wave Born approximation~\cite{khakoo2003electron}, relativistic-coupled cluster~\cite{vcivzek1966correlation,bharti2019application}, and optical potential modeling using density functional theory~\cite{salvat2005elsepa} incorporated electron exchange and correlation effects. Thus, high-accurate \textit{e}-atom scattering predictions were possible, including routinely prevalent diffraction structures matching measurements~\cite{predojevic2007elastic, williams1978electron, brown2003scattering, milisavljevic2005differential, marinkovic2023cross, jensen1978elastic, wang1994cross, fregeau1956elastic, lewis1974elastic_ar, gibson1998low, williams1975scattering, gianturco1994elastic,lewis1974elastic, williams1975scattering, gianturco1994elastic, lewis1974elastic, williams1975scattering, lewis1974elastic}. 


However, the electron diffraction from atoms is largely counterintuitive, since the localized orbital-density under long-range interaction monotonically fizzles away with the distance from the nucleus and thus does not produce a ready ``edge". In fact, no clear attempt has been made to interpret the \textit{e}-atom diffraction, except simply assuming that the atom presents a ``dark" disk~\cite{cao2018}. We attempt to fill this gap in the current theory-experiment study. This new understanding ushers a valuable benefit: by transferring the diffraction pattern in the elastically scattered angular distribution to the Fourier conjugate space for various incident momenta it is possible to image the effective atomic scattering potential.
\begin{figure*}[t]
\includegraphics[width=17 cm]{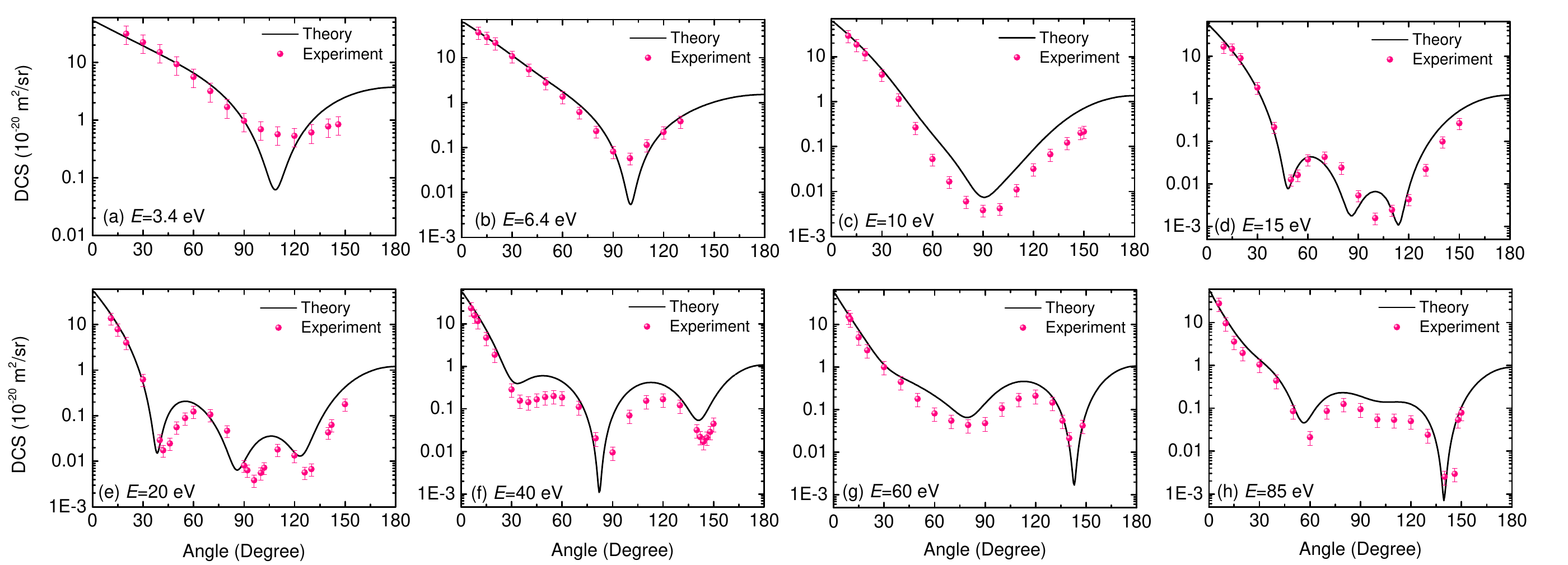}
\caption{Absolute angular DCS of \textit{e}-Cd for collision energies 3.4 (a), 6.4 (b), 10 (c), 15 (d), 20 (e), 40 (f), 60 (g), and 85 (h) eV are compared for current measurement versus calculation. The TCS comparison in Fig.\,S2 is given in SM~\cite{SM_e-atomsc}. \label{fig1}}
\end{figure*}   

Diffraction is also commonplace in the photoionization (PI) of symmetric molecules~\cite{azizi2024} and clusters~\cite{rudel2002imaging,magrakvelidze2015}. Even though for the PI of a spherical system the angular distribution mode of diffraction is forbidden due to the symmetry, the diffraction is accessible in the photoelectron energy distribution. In distinction, for the scattering by even a spherical system, the presence of the projectile breaks the overall symmetry. In any case, a simpler way to understand the diffraction in PI is to consider the spatial gradient of the ground state potential that the photoelectron escapes. This gradient is stronger at the edge of a delocalized electron distribution and at the interatomic bond edges in molecules. Hence, local electron emissions occur from these regions due to the availability of strong ionizing force (potential gradient), the key quantity within the acceleration gauge formalism of PI. Consequently, diffraction results. In contrast, for a single atom, Coulomb potential entirely forbids diffraction in PI, since $\frac{d}{dr}(-\frac{Z}{r})$ being continuous suggests no cite-specific emission. So the fundamental question is: what is the leading process that underpins diffraction in scattering from atoms, which do not have density edges? To elucidate, we employ an interpretative model that reveals a centroidal diffraction mechanism. This effectuates average diffractor sizes as a function of $E$ that map the net profile of the scattering potential. 

Details of the theory are given in Supplementary Material (SM)~\cite{SM_e-atomsc}. The electron density of the atomic target is obtained using the self-consistent relativistic Dirac-Fock formalism~\cite{johnson2007atomic}. The \textit{e}-atom interaction potential is constructed in an optical model approach~\cite{salvat2005elsepa}, which incorporates effects of exchange and correlation. Scattering parameters are computed within the relativistic partial wave framework. The scattering phase shifts, $\delta_\kappa$, are obtained by numerically solving the coupled radial Dirac equations with the interaction potential $V_{\mbox{\scriptsize Opt}}$~\cite{rose1961relativistic} as  
\begin{eqnarray}
       \frac{dP_{E\kappa}(r)}{dr} &=& -\frac{\kappa}{r}P_{E\kappa}(r)+\frac{E-V_{\mbox{\scriptsize Opt}}+2m_e c^2
    }{c\hbar} Q_{E\kappa}(r)\\
     \frac{dQ_{E\kappa}(r)}{dr} &=& -\frac{E-V_{\mbox{\scriptsize Opt}}}{c\hbar}P_{E\kappa}(r)+\frac{\kappa}{r}Q_{E\kappa}(r),
\end{eqnarray}
where $P_{E\kappa}(r)$ and $Q_{E\kappa}(r)$ represent the large and small component of the radial wave function, respectively; see the meaning of other terms in SM~\cite{SM_e-atomsc}.\nocite{furness1973semiphenomenological, lide2004crc, salvat2003optical,  perdew1981self, salvat1995accurate, press2002art} Using $\delta_\kappa$, the scattering amplitudes are calculated, from which the differential cross section (DCS) is obtained as ~\cite{wavemech_applications}: 
\begin{eqnarray}\label{dcs}
\frac{d\sigma}{d\Omega} &=& |f(k,\theta)|^2+|g(k,\theta)|^2,
\end{eqnarray}
where $f(k,\theta)$ and $g(k,\theta)$ are direct and spin-flip amplitudes. Previous studies using Dirac partial wave analysis showed excellent agreement between calculated and experimental cross sections for atoms~\cite{ismail2016elastic,tripathi2000spin,adibzadeh2004elastic,hussain2022differential}. The DCS depends on the magnitude of incident momentum $k=\sqrt{2E}$, in atomic units (a.u.), and scattering angle $\theta$. In elastic scattering, the magnitude of momentum transfer 
\begin{eqnarray}\label{tran-mom}
q=2k\sin (\theta/2).
\end{eqnarray}
Obviously, the variation in $q$ can be affected by either scanning $\theta$ with fixed $E$ ($k$) or tuning $E$ while looking along a fixed $\theta$, laying out a 2D diffraction landscape~\cite{aiswarya2024simultaneous}. 

Experimental data for \textit{e}-Cd collision are collected using a crossed-beam spectrometer where a mono-energetic electron beam is focused onto the atomic beam, and the scattered electrons are detected as a function of $E$ and $\theta$. The principal parts of the spectrometer are: (i) the vacuum chamber containing an oven for metals, an electron source with a monochromator and a scattered electron detector with an analyzer; (ii) the system of vacuum pumps and electric power supplies; and (iii) measuring instruments~\cite{filipovic1988, milisavljevic2004, rabasovic2008}. The oven heated by coaxial wires is used to produce a gas-phase atoms beam by vaporizing a solid sample~\cite{marinkovic2007}. The monochromator and analyzer were constructed as systems of cylindrical electrostatic lenses with hemispherical dispersion elements. The electron optics was designed along principles outlined by Kuyatt~\cite{kuyatt1967} and by Chutjian~\cite{chutjian1974}. 
The $E$ range of the impact electrons was from several eV to 100 eV and $\theta$ ranged from 6$^\circ$ to 150$^\circ$. The achieved energy resolution of energy-loss spectra was 50 meV, while the angular resolution of the spectrometer was 1.5$^\circ$ in DCS measurements.


In Figure \ref{fig1}, our measured angular DCS in the absolute scale (numerical data in SM~\cite{SM_e-atomsc}) of elastic \textit{e}-Cd for a selection of $E$ compares well with our calculations. Results show maxima and minima, in varied intensities, representing bright and dark diffraction fringes. Even though at lower $E$ theory somewhat over-sharpens the minima, the qualitative agreement is obvious. Generally, this fringe pattern is seen to shrink with increasing $E$. 
\begin{figure} [h]
\includegraphics[width=9 cm]{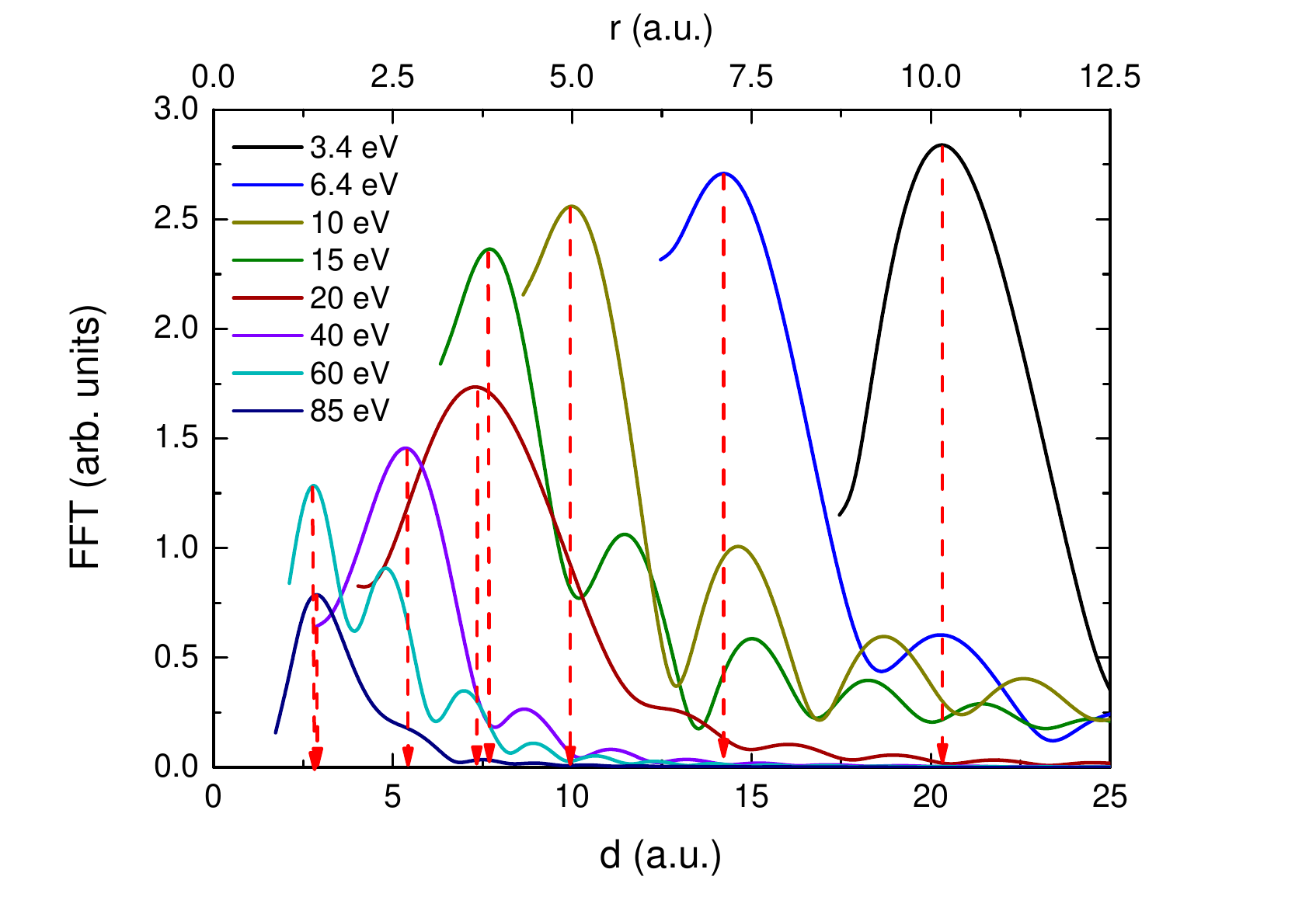}
\caption{FFT spectra of calculated DCS for collision energies in Fig.\,\ref{fig1}. Arrows point the peak locations.   \label{fig2}}
\end{figure}   

The $\theta$ scan of the computed DCS for a given $k$ can be converted to $q$-scale by \eq{tran-mom}. As characteristic of the diffraction, the FT of the pattern in $q$ yields the diffraction length in the reciprocal (real) space that connects to the size of the diffractor electrons experience. We carry out FT following Ref.\ \cite{aiswarya2024simultaneous,hervieux2017ubiquitous,rudel2002imaging,mccune2008unique}; a brief description of the method is included in SM~\cite{SM_e-atomsc}. To capture the pure diffraction component in DCS, we curve-fit the non-diffractive background~\cite{mccune2008unique} and subtract it from the full signal before applying a fast FT (FFT). Resulting FFT spectra, for the set of $E$ in Fig.\ \ref{fig1}, are presented in Figure \ref{fig2}.  These spectra emerge in two-times-radial scale $d=2r$ (see in the following). The FFT reveals distinct peaks for different $E$. 
As evident, the peak location, at 20.2 a.u.\ for the lowest 3.4 eV, gradually decreases to 2.82 a.u.\ for the highest 85 eV. Thus, as $E$ increases, the FFT peaks move closer to the nucleus, indicating that higher the electron energy the smaller the diffraction length it senses. So, the natural question is: What makes the progressively energetic electron to see smaller diffracting size? 
\begin{figure} [t]
\includegraphics[width=7 cm]{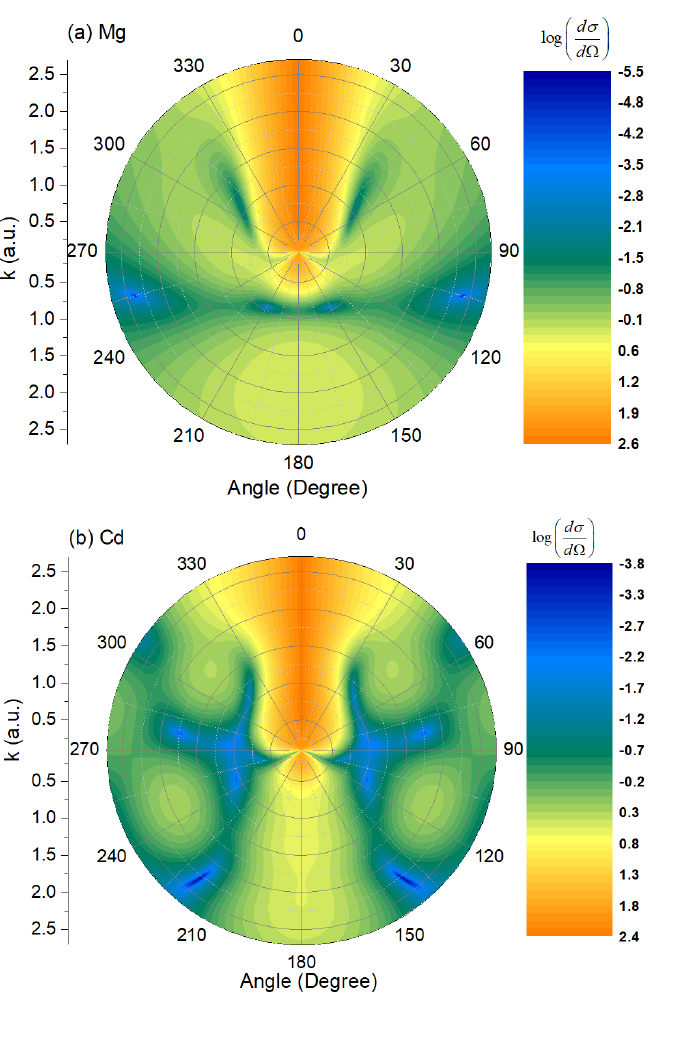}
\caption{Diffractograms calculated for (a) \textit{e}-Mg and (b) \textit{e}-Cd.   \label{fig3}}
\end{figure}   

Before we answer this question, let us illustrate by iso-surface images in Figure \ref{fig3}, the complete 2D diffractrograms computationally simulated for a smaller Mg and a larger Cd. Both modes, $\theta$-scan at fixed $k$ and $k$-scan at fixed $\theta$, are depicted in polar graphs of DCS. $k$ is plotted along the radius, such that at a given radius going over a circle is the former mode, while at a given direction going along a radius is the latter. As expected, the maximum DCS is noted in the forward direction. More importantly, the overall structures being richer for Cd points a shrunk pattern for a larger target from the inverse relation of the diffraction length and inter-fringe separation.

We introduce a model pathology to understand the evolution of the FT-derived diffracting length as $E$ ($k$) tunes. Consider the elastic Born amplitude of in a.u.,
\begin{eqnarray}\label{born-amp}
    f_{\mbox{\scriptsize B}} (k,\theta) &=& 2\int_0^\infty dr r^2 \frac{\sin(qr)}{qr}V(r).
\end{eqnarray}
Recalling \eq{tran-mom}, we may write
\begin{eqnarray}\label{q-in-b}
    q &\approx& 2k - k\cot^2(\theta/2).
\end{eqnarray}
In approximating \eq{q-in-b} we ignore the forward scattering, where $\cot(\theta/2)$ is large, so we retain up to the linear power of $\cot^2(\theta/2)$ in the Taylor expansion. This goes without any loss of generality, since no appreciable diffraction exists in the forward direction (Fig.\,3). Plugging \eq{q-in-b} in the relevant part of \eq{born-amp} one obtains,
\begin{eqnarray}\label{born-b}
 f_{\mbox{\scriptsize B}} (k,\theta) &\approx& \frac{2}{q} \int_0^\infty dr r \sin[2kr-kr\cot^2(\theta/2)]V(r). 
\end{eqnarray}
 It is now helpful to split the amplitude between a non-diffractive steady background (bg) and a pure diffractive (df) component, and recognize that the dominant diffraction arises from $r=b$. Here, $b = (L/k^2)\cot(\theta/2)$ is the classical impact parameter of collision with $L=Z/(4\pi\epsilon_0)$. This enables a further approximation of the diffraction amplitude as,
 \begin{eqnarray}\label{born-b-df}
 f_{\mbox{\scriptsize B}}^{\mbox{\scriptsize (df)}} (k,\theta) &\approx&  \frac{2}{q}b(\theta)\sin[2kb(\theta)]V[b(\theta)],
 \end{eqnarray}
 in which, obviously, $\theta$-scan maps into the variation of $b$, with $b$ decreasing towards the backward scattering.
\begin{figure*} [!htb]
\centering
\includegraphics[width=7.1in]{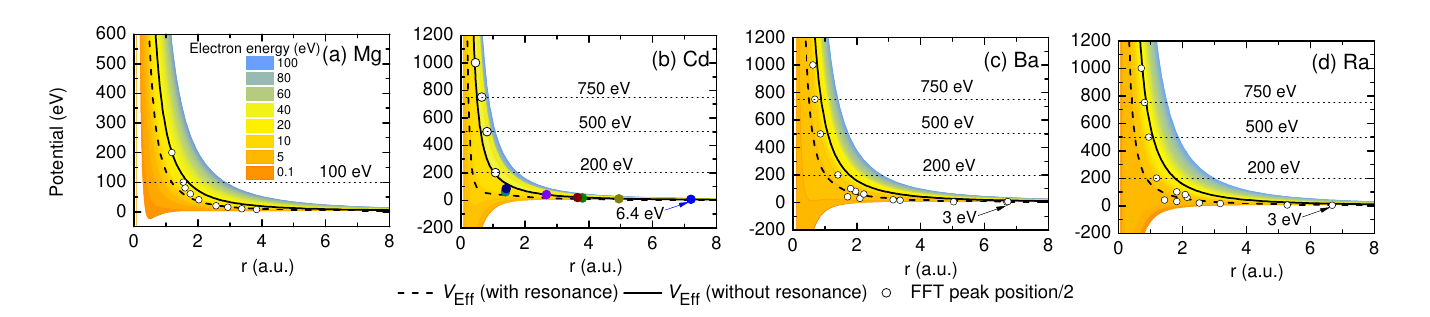} 
\caption{Effective potential $V_{\mbox{\scriptsize Eff}}$ [Eq.\,(\ref{V_avg})] averaged over the electron energies ($E_i$) in the range 0.1-100 eV using TCS with and without the resonance. The color bands correspond to $V(r,E_i)$. FFT peak locations, divided by two, for different collision energies are plotted by dots. For Cd (b), the dots are color-matched to respective curves in Fig.\,\ref{fig2}. Collision energies of some dots are indicated by horizontal lines and a few arrows. \label{fig4}}
\end{figure*}   

 For clarity, we invoke average $b$ as $b_{\mbox{\scriptsize c}}$ to write $b=b_{\mbox{\scriptsize c}} + \alpha$, where $\alpha >$ and $< 0$ define outer and inner locations of $b$ from $b_{\mbox{\scriptsize c}}$ directly varying with $\theta$. Hence, \eq{born-b-df} reads as,
 \begin{eqnarray}\label{born-b-df2}
 f_{\mbox{\scriptsize B}}^{\mbox{\scriptsize (df)}} (k,\theta) &\approx&  \sin[2k(b_{\mbox{\scriptsize c}}+ \alpha)] \left( \frac{2}{q}(b_{\mbox{\scriptsize c}}+ \alpha) V(b_{\mbox{\scriptsize c}}+ \alpha) \right) \nonumber \\
 &=& \sin[2kb_{\mbox{\scriptsize c}}+ 2k\alpha] \mathcal{F}(\alpha),
 \end{eqnarray}
 such that we express the full amplitude in the form:
 \begin{eqnarray}
  f_{\mbox{\scriptsize B}} (k,\theta) &\approx&  f_{\mbox{\scriptsize B}}^{\mbox{\scriptsize (bg)}} + \sin(2kb_{\mbox{\scriptsize c}}+ 2k\alpha) \mathcal{F}(\alpha).
 \end{eqnarray}
 \eq{dcs}, ignoring $|\mathcal{F}|^2$, yields the DCS as
 \begin{eqnarray}\label{dcs2}
  \frac{d\sigma}{d\Omega}_{\mbox{\scriptsize B}}  &\sim&   \frac{d\sigma}{d\Omega}_{\mbox{\scriptsize B}}^{\mbox{\scriptsize (bg)}} + 2\sin[2kb_{\mbox{\scriptsize c}}+ 2k\alpha] f_{\mbox{\scriptsize B}}^{\mbox{\scriptsize (bg)}} \otimes \mathcal{F}(\alpha).
 \end{eqnarray}
We argue, the interference term from quantum coherence containing $\otimes$ above embodies a notion of diffraction. 

Evidently in \eq{dcs2}, the diffraction dominates with no dephasing as $\alpha=0$ or $b=b_{\mbox{\scriptsize c}}$. Here, the de\,Broglie wavelength of $k$ will match $2b_{\mbox{\scriptsize c}}$, rendering $2b_{\mbox{\scriptsize c}}$ to be the diffraction length. Including $\alpha$'s, however, $2b_{\mbox{\scriptsize c}}$ modifies to a {\em centroid} location due to the averaging, weighted by $f_{\mbox{\scriptsize B}}^{\mbox{\scriptsize (bg)}} \otimes \mathcal{F}(\alpha)$ in \eq{dcs2} for individual Fourier components. Thus, the position of the FT peak provides a centroidal diffraction-size for a given $k$ ($E$). We show below, by varying $E$ and applying FT to angular DCS we map out a profile that images the scattering potential. 

Since the scattering potential is a function of $E_i$, we employ the weighted averaging (SM~\cite{SM_e-atomsc} for details)
\begin{equation}\label{V_avg}
    V_{\mbox{\scriptsize Eff}}(r)=\frac{\sum_{i}V(r,E_i)\sqrt{\sigma(E_i)}}{\sum_{i}\sqrt{\sigma(E_i)}}
\end{equation}
to determine the effective scattering potential the projectile explores. As the diffraction vanishes in total (angle-integrated) CS (TCS), $\sigma$, such that $\sqrt{\sigma} \sim |f_{\mbox{\scriptsize B}}^{\mbox{\scriptsize (bg)}}|$, the weight factor in \eq{V_avg} is consistent with $f_{\mbox{\scriptsize B}}^{\mbox{\scriptsize (bg)}} \otimes \mathcal{F}(\alpha)$ in \eq{dcs2} for each Fourier component. Indeed, remarkably, the half of the FFT peak positions as $E$ varies, like in Fig.\,\ref{fig2}, closely follow the $V_{\mbox{\scriptsize Eff}}$ shape for Cd, panel (b), in Figure \ref{fig4}. We also include results to verify this prediction for Mg, Ba and Ra which reinforces the generality of the phenomenon, at least for closed shell atoms/ions. Note, for the smallest atom Mg, we do not find enough diffraction signal for FFT at $E=$ 500, 750 and 1000 eV. $V(r,E_i)$ over the range 0.1-100 eV considered for averaging are shown as color-maps in Fig.\,\ref{fig4}. In essence, the electron with $E$ probes a range of $V(r,E_i)$ in $\theta$-differential scattering. The dominant diffraction from $b_{\mbox{\scriptsize c}}$, varying with $\theta$, yields a centroid diffraction-length, which is automatically $r$ of $V_{\mbox{\scriptsize Eff}} = E$.

We make two observations in Fig.\,\ref{fig4}. (i) The low-$E$ end of potential bands exhibits attractive (negative) shapes which becomes stronger with larger targets. These spawn resonances in TCS at low $E$ from metastable binding of slow projectiles (Fig.\,S1 in SM~\cite{SM_e-atomsc}). One effect of these resonances is the accumulation of some FFT data points for Cd, Ba and Ra that fall outside the trend lines. (ii) Assuming the resonances not participating in the potential imaging, we applied non-resonant $\sigma$'s in \eq{V_avg} to plot. However, we also plot the potential weighted with resonant $\sigma$'s. As seen, while the latter show better propensity to match with the FFT trend at lower $E$, the former succeeds at higher $E$. There is another implication of this result. The sharper resonances suggest longer lifetimes of slow electron's binding. Thus, for a diffraction study of scattering time delay~\cite{aiswarya2024delay}, this longer delay trapped in resonance states may remain insensitive to fast scattering in imaging scattering potential~\cite{Ca-exception}.  

Imaging the scattering potential through {\em e}-atom diffraction promises impacts in applied fields~\cite{daimon2024,ji2023,harilal2022,rosandi2024}. 
In particular, they may have large-scale modeling implications. Adopting the complex energy-dependent potentials is difficult and expensive where one needs to recalculate potentials for each local configuration impeding efficiency in multi‐scale simulations, say, in Boltzmann transport or Monte Carlo~\cite{madsen2021abtem,geant4_msc,salvat2011penelope}. A universal potential, fairly accurate for each atomic species, simplifies the method. Thus, the average elastic‐scattering potential for a single atom may enable building models of electron-beam spectroscopy~\cite{jablonski2000effects} in materials and even extreme states in plasma or warm dense matter~\cite{johnson2012thomson}. This may also support models of beams interacting with many atoms, of image contrast in electron microscopy~\cite{rez1982elastic,treacy2011z}, and of radiation damage. The cost of deriving highly accurate potentials, namely, via coupled‐cluster methods~\cite{sahoo2022constructing}, is tall. The trade-off between accuracy and speed matters.

To conclude, this theory-experiment joint study presents and compares absolute-scale angularly-distributed DCS for \textit{e}-Cd elastic scattering for a range of impact energies. While the diffraction patterns in \textit{e}-atom scattering is known for a long time, the understanding of the mechanism was vastly elusive. We perform a fully quantum calculation but illustrate an approach to uncover a centroidal diffraction phenomenon intrinsic to scattering from edge-less, smudgy atomic boundaries, as opposed to the regular diffraction from semi-discrete boundary of molecular potentials and bonds. The approach demonstrates that the overall diffracting length, obtained from the FT of DCS fringes to the reciprocal space, as a function of energy can delineate the scattering potential -- a cardinal collision parameter not directly observable in experiments. Proving the universality of the result, the method is shown to be also successful for Mg, Ba, and Ra. This offers a spectroscopic route to capture a quantitative scattering force within the long-range atomic potential and opens a new route to explore atomic interactions at a fundamental level. The study may develop technology and research applications, and may infuse scopes in the field of ultrafast electron diffraction~\cite{williamson1992,centurian2022}.

\begin{acknowledgements} 
 The research is supported by: CRG/2022/000191, India (J.J.); Institute of Physics Belgrade, through a grant from the Ministry of Science, Technological Development and Innovation of the Republic of Serbia (N.S.); Science Fund of the Republic of Serbia, grant No. 6821-ATMOLCOL (B.P.M); US National Science Foundation Garnt Nos.\ PHY-2110318 and PHY-2512850 (H.S.C).
\end{acknowledgements}


\bibliography{reference-main}

\end{document}


\title{Supplementary material for ``Imaging scattering potential in centroidal atomic diffraction of elastic electrons''}

\author{R. Aiswarya}
\affiliation{Department of Physics, Indian Institute of Technology Patna, Bihar 801103, India}

\author{Jobin Jose}
\email[]{jobin.jose@iitp.ac.in}
\affiliation{Department of Physics, Indian Institute of Technology Patna, Bihar 801103, India}

\author{Nenad Simonovi\'{c}}
\affiliation{Institute of Physics Belgrade, Pregrevica 118, Belgrade, Serbia}

\author{Bratislav P. Marinkovi\'{c}}
\email[]{bratislav.marinkovic@ipb.ac.rs}
\affiliation{Institute of Physics Belgrade, Pregrevica 118, Belgrade, Serbia}

\author{Himadri S. Chakraborty}
\email[]{himadri@nwmissouri.edu}
\affiliation{School of Natural Sciences, D.L.\ Hubbard Center for Innovation, Northwest Missouri State University, Maryville, Missouri 64468, USA}

\maketitle

\section{Interaction Potential details} \label{Pot_details}
In this study, the optical potential model of atomic targets combined with Dirac partial wave analysis is employed. The total interaction potential in \textit{e}-atom scattering can generally be expressed by means of an optical-model potential \cite{salvat2005elsepa}:
\begin{equation}
\label{V_pot}
    V_{\mbox{\scriptsize Opt}}(r,E)=V_{\mbox{\scriptsize Static}}(r)+V_{\mbox{\scriptsize Ex}}(r,E)+V_{\mbox{\scriptsize CP}}(r)+iV_{\mbox{\scriptsize Abs}}(r,E),
\end{equation}
where $V_{\mbox{\scriptsize Static}}$ represents electrostatic interaction potential, $V_{\mbox{\scriptsize Ex}}$ is the exchange potential, $V_{\mbox{\scriptsize CP}}$ is the correlation-polarization potential, and $V_{\mbox{\scriptsize Abs}}$ denotes the imaginary absorption potential. $V_{\mbox{\scriptsize Abs}}$ is also considered to account for the finite probability of inelastic processes, although its effect on the scattering cross section becomes significant only at 1 MeV and higher projectile energies \cite{salvat2005elsepa}. 

$V_{\mbox{\scriptsize Static}}$ includes contributions from the \textit{e}-\textit{e} and \textit{e}-nuclear interaction. At a position \textit{r}, $V_{\mbox{\scriptsize Static}}$ can be calculated using the following expression:
\begin{multline}
\label{Static_pot}
    V_{\mbox{\scriptsize Static}}(r)=-\frac{Ze^2}{r} \\
    - \left( \frac{e}{r}\int_0^r 4\pi\rho_e(r')r'^2dr' +\int_r^\infty 4\pi\rho_e(r')r'dr' \right),
\end{multline}
where $Z$ is the atomic number of the target atom and $\rho_e(r)$ is its electronic charge density. In the present study, the electron densities of atoms are computed using the Dirac-Fock (DF) methodology \cite{salvat2005elsepa, johnson2007atomic}. $V_{\mbox{\scriptsize EX}}$ is formulated following the work of Furness and McCarthy \cite{furness1973semiphenomenological} as, 
\begin{multline}
    V_{\mbox{\scriptsize EX}}(r,E)=\frac{1}{2}[E-V_{\mbox{\scriptsize Static}}(r)]\\
    -\frac{1}{2}\left[(E-V_{\mbox {\scriptsize Static}}(r))^2+4\pi a_0 e^4 \rho_e(r)\right]^{1/2}.
\end{multline}
Here $a_0$ is Bohr radius and $E$ the impact energy of the incident electron. This non-local exchange potential form can be derived from the WKB-like approximation of the wavefunction. 

We employ a correlation–polarization potential model that combines the long-range Buckingham polarization potential with the correlation potential based on the local-density approximation (LDA). The combined potential is expressed as \cite{salvat2005elsepa}:
\begin{equation}
    V_{\mbox{\scriptsize CP}}(r)=\begin{cases}
    \max[V_{\mbox{\scriptsize Co}}(r),V_{\mbox{\scriptsize Pol}}(r)] & if \quad r<r_{\mbox{\scriptsize cp}}\\
    V_{\mbox{\scriptsize Pol}}(r) & if \quad r\ge r_{\mbox{\scriptsize cp}}
    \end{cases},
\end{equation}
where $r_{\mbox{\scriptsize cp}}$ is the outer radius at which the correlation potential, $V_{\mbox{\scriptsize Co}}(r)$, and the polarization potential, $V_{\mbox{\scriptsize Pol}}(r)$, merge. The long-range $V_{\mbox{\scriptsize Pol}}$ is modeled using the Buckingham potential form \cite{salvat2005elsepa}:
\begin{equation}
\label{pol_pot}
    V_{\mbox{\scriptsize Pol}}=\frac{-\alpha_d e^2}{2\left( r^2+d^2\right)^2}. 
\end{equation}
Here $\alpha_d$ is the dipole polarizability of the target atom \cite{lide2004crc} and $d$ is the cut-off parameter, which avoids divergence at $r=0$. The values of $\alpha_d$ are taken from the experimental data \cite{lide2004crc}. The cut-off parameter is defined as \cite{salvat2003optical}: 
\begin{equation}
    d^4=\frac{1}{2} \alpha_d a_0 Z^{-1/3} b^2_{\mbox{\scriptsize Pol}},
\end{equation}
where $b_{\mbox{\scriptsize Pol}}$ is an energy-dependent adjustable parameter which is obtained by fitting the DCS at small angles \cite{salvat2003optical}. $V_{\mbox{\scriptsize Co}}$ is obtained from LDA, by approximating that the projectile at \textit{r} has the same correlation energy as if it were moving in a free electron gas with a density equal to the local atomic electron density $\rho_e(r)$ \cite{salvat2005elsepa}. For defining $V_{\mbox{\scriptsize Co}}$, a density parameter, $r_s$ is introduced: 
\begin{equation}
    r_s\equiv \frac{1}{a_0}\left[\frac{3}{4\pi\rho_e(r)}\right]^{1/3},
\end{equation}
which, expressed in units of the Bohr radius, $a_0$, is the radius of the sphere that, on average, contains one electron of the gas. For electrons,$V_{\mbox{\scriptsize Co}}$  is parameterized as suggested by Perdew and Zunger as \cite{perdew1981self}:\\ 
\\
For $r_s<1$,
\begin{multline}
    V_{\mbox{\scriptsize Co}}(r)=-\frac{e^2}{a_0}(0.0311 \ln(r_s)-0.0584\\
    +0.00133r_s \ln(r_s)-0.008 4r_s).
\end{multline}
For $r_s\ge1$,
\begin{equation}
    V_{\mbox{\scriptsize Co}}(r)=-\frac{e^2}{a_0}\beta_0\frac{1+(7/6)\beta_1r_s^{1/2}+(4/3)\beta_2r_s}{(1+\beta_1r_s^{1/2}+\beta_2r_s)^2}.
\end{equation}
Here $\beta_0$=0.1423, $\beta_1$=1.0529 and $\beta_2$=0.3334. The correlation-polarization potential correction to the differential cross section (DCS) is appreciable only for slow projectiles and at small scattering angles \cite{salvat2005elsepa}. 

The $V_{\mbox{\scriptsize Abs}}$ form used in the current study is the same suggested by Salvat \cite{salvat2003optical}:
\begin{multline}
    V_{\mbox{\scriptsize Abs}}(r,E)=\sqrt{\frac{2(E_L+m_ec^2)^2}{m_ec^2(E_L+2m_e c^2)}}\\
    \times A_{\mbox{\scriptsize Abs}}\frac{\hbar}{2}[\nu_L \rho_e(r) \sigma_{bc}(E_L,\rho_e,\Delta)],
\end{multline}
where $m_e $ is the the electron rest mass and $E_L$ is the local kinetic energy defined for the incident energy $E$ as $E_L(r)=E-V_{\mbox{\scriptsize Static}}(r)-V_{\mbox{\scriptsize Ex}}(r)$. The local velocity, $\nu_L=(2E_L/m_e)^{1/2}$, describes the projectile interaction as if it were moving within a homogeneous gas of electron density $\rho_e$. $\sigma_{bc}(E_L,\rho_e,\Delta)$ is  the cross section when collision energy transfer is grater than energy gap $\Delta$ (the first excitation energy of the target atom). The empirical parameter $A_{\mbox{\scriptsize Abs}}$ is set to 2 in our calculations, as recommended by Salvat based on an analysis of the DCS in previous studies \cite{salvat2005elsepa, salvat2003optical}.

While the calculations are carried out with the full potential $V_{\mbox{\scriptsize Opt}}$, we now describe a closely approximated version of the potential used to interpret our main results. When dealing with the particle collision in a central potential, a repulsive effect naturally arises due to the centrifugal term $V_{\ell}=\ell(\ell+1)/r^2$, which repels the particle from the origin. Since the total cross section (TCS) [Eq. (\ref{tcs})] is basically the sum of different individual partial cross sections (PCS), we have approximately defined an average centrifugal potential, $V_{CF}$, of the form:
\begin{equation}
\label{V_cf}
    V_{\mbox{\scriptsize CF}}=\frac{\displaystyle \sum_{\ell}\sigma_{\ell}*V_{\ell}}{\displaystyle \sum_{\ell}\sigma_{\ell}}.
\end{equation}
Thus, a working version of the interaction potential, $V$, can be composed by combining $V_{\mbox{\scriptsize CF}}$ [Eq. (\ref{V_cf})] and $V_{\mbox{\scriptsize Opt}}(r,E)$ [Eq. (\ref{V_pot})] as: 
\begin{equation}
\label{V_total}
    V(r, E)=V_{\mbox{\scriptsize Opt}}(r,E)+V_{\mbox{\scriptsize CF}}(r)-iV_{Abs}(r,E).
\end{equation}
In $V$, the absorption potential is excluded as it only becomes significant at very high energies. Additionally, to compute $V_{\mbox{\scriptsize CF}}$ partial waves up to $\ell=20$ are considered, since contributions from higher partial waves are found negligible within the projectile energy range of the study. Obviously, $V$ varies with the incident energy \cite{salvat2005elsepa}. To further aid the analysis, we define an energy-average scattering potential, $V_{\mbox{\scriptsize Eff}}$, as, 
\begin{equation}\label{V_avg}
    V_{\mbox{\scriptsize Eff}}(r)=\frac{\sum_{i}V(r,E_i)\sqrt{\sigma(E_i)}}{\sum_{i}\sqrt{\sigma(E_i)}},
\end{equation}
where $\sigma$ is the TCS.
\section{Cross section calculation}

The dynamics of the relativistic projectile electron interacting with the atomic target through the spherically symmetric potential $V_{\mbox{\scriptsize Opt}}(r,E)$ [Eq. (\ref{V_pot})] is governed by the Dirac equation \cite{rose1961relativistic}:
\begin{equation}
\label{DF-eq}
    \left[ c\vec{\alpha}.\vec{p}+\beta m_e c^2 + V_{\mbox{\scriptsize Opt}}(r,E)\right]\psi_{E\kappa m}(\vec{r})=E \psi_{E\kappa m}(\vec{r}).
\end{equation}
Here $\vec{\alpha}$ and $\beta$ are the $4 \times 4$ Dirac matrices. The momentum vector $\vec{p}=\hbar k$ with \textit{k} being the relativistic wave number of projectile electron and \textit{c} is the velocity of light in vacuum. The relativistic quantum number $\kappa=(\ell-j)(2j+1)$, where $\ell$ and $j$ are the orbital and total angular momentum quantum numbers. The values  of $\ell$ and $j$ can be obtained from $\kappa$ value as $j=|\kappa|-1/2$ and $\ell=j+\kappa/(2|\kappa|)$. The solution to the Eq. (\ref{DF-eq}) can be expressed as \cite{rose1961relativistic}: 
\begin{equation}
    \psi_{E\kappa m} (\vec{r})=\frac{1}{r} \left( \begin{array}{c}
       P_{E\kappa}(r)\Omega_{\kappa,m}(\hat{r})   \\
       iQ_{E\kappa}(r)\Omega_{-\kappa,m}(\hat{r})   
    \end{array} \right),
\end{equation}
where $P_{E\kappa}(r)$ and $Q_{E\kappa}(r)$ represent the large and small component of the radial wave function respectively and $\Omega_{\kappa,m}(\hat{r})$ are the spherical spinors. $P_{E\kappa}(r)$ and $Q_{E\kappa}(r)$ satisfy the following coupled differential equation \cite{rose1961relativistic}:  

\begin{multline}
    \frac{dP_{E\kappa}(r)}{dr}=-\frac{\kappa}{r}P_{E\kappa}(r)+\frac{E-V_{\mbox{\scriptsize Opt}}(r,E)+2m_e c^2
    }{c\hbar} Q_{E\kappa}(r),
\end{multline}

\begin{equation}
     \frac{dQ_{E\kappa}(r)}{dr}=-\frac{E-V_{\mbox{\scriptsize Opt}}(r,E)}{c\hbar}P_{E\kappa}(r)+\frac{\kappa}{r}Q_{E\kappa}(r).
\end{equation}
The above coupled equations are solved numerically using the subroutine package RADIAL \cite{salvat1995accurate} to obtain the relativistic scattering phase shift $\delta_\kappa$. The radial component of wave function, $P_{E\kappa}(r)$, is normalized to give unit amplitude oscillations in the asymptotic region. For finite range fields, $P_{E\kappa}(r)$ in the asymptotic region can be approximated as:  
\begin{equation}
     P_{E\kappa}(r) \simeq \sin \left( kr-\frac{\ell\pi}{2}+\delta_\kappa \right).
\end{equation}

In relativistic case, for a given $\kappa$, two values of $\ell$ are possible. Consequently two phase shift are obtained: $\delta_{\kappa=-\ell-1}$ and $\delta_{\kappa=\ell}$. Additionally, there will be two scattering amplitude corresponding to the direct scattering, $f(k,\theta)$, and the spin-flip scattering, $g(k,\theta)$. The scattering amplitudes thus take the form \cite{wavemech_applications, mott1933theory}: 
\begin{multline}
\label{direct_term}
    f(k,\theta)=\frac{1}{2ik}\sum_{\ell=0}^{\infty}\{(\ell+1)[exp(2i\delta_{\kappa=-\ell-1})-1]\\
    +\ell[exp(2i\delta_{\kappa=\ell})-1]\}P_{\ell}(\cos\theta),
\end{multline}
\smallskip  
\begin{multline}
\label{spin_flip_term}
    g(k,\theta)=\frac{1}{2ik}\sum_{\ell=0}^{\infty}\{exp(2i\delta_{\kappa=\ell})\\
    -exp(2i\delta_{\kappa=-\ell-1})\}P^{1}_{\ell}(\cos\theta).
\end{multline}
Here, $P_{\ell}(\cos\theta)$ and $P^{1}_{\ell}(\cos\theta)$ represent Legendre polynomial and associated Legendre polynomial of order $\ell$. Utilizing the above expressions for the scattering amplitudes, the DCS can be written as \cite{wavemech_applications}:
\begin{equation}
\label{dcs}
\frac{d\sigma}{d\Omega} = |f(k,\theta)|^2+|g(k,\theta)|^2.
\end{equation}
\renewcommand{\thefigure}{S1}
\begin{figure} [t] 
\includegraphics[width=8.2 cm]{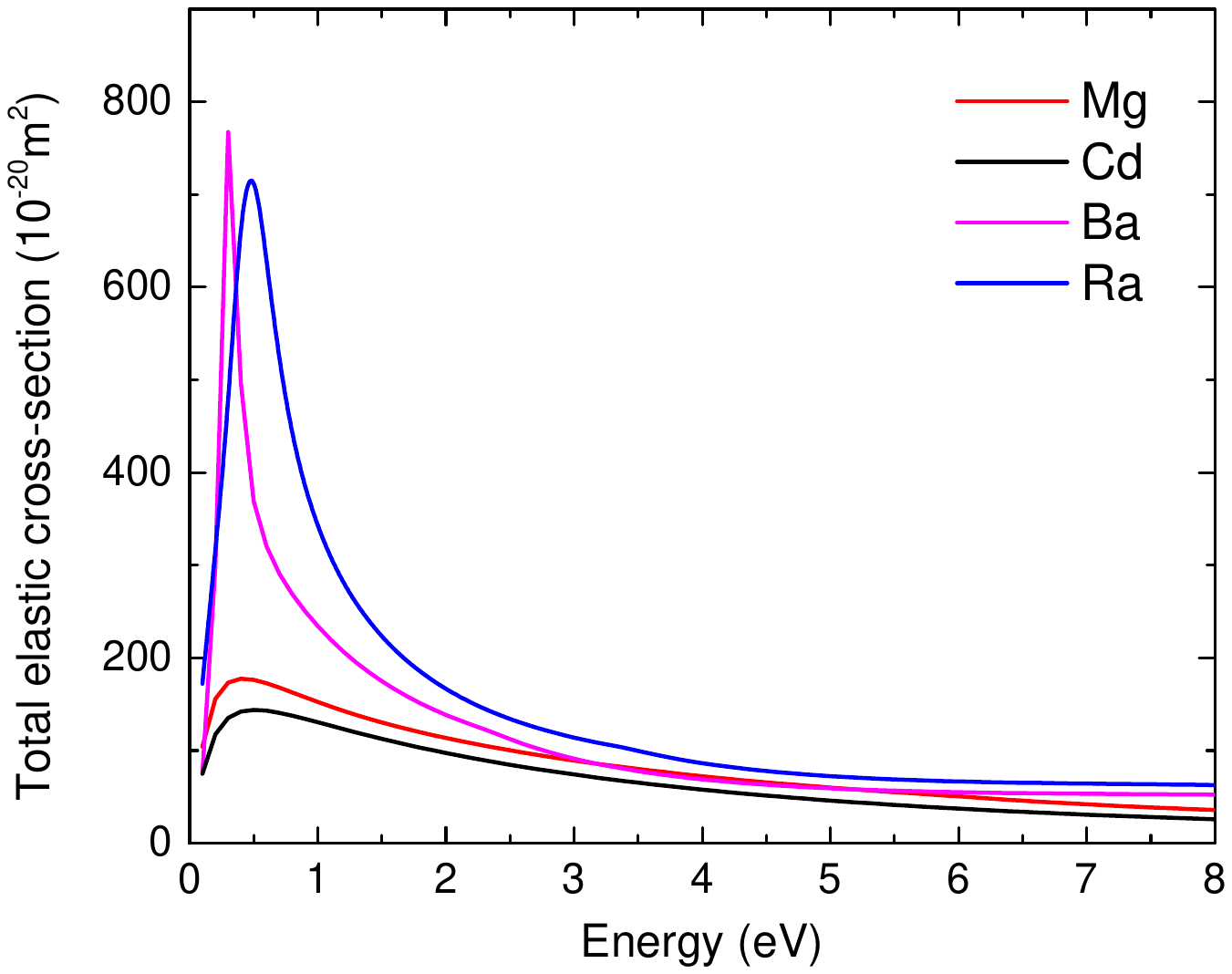}
\caption{TCS of Mg (red solid line), Cd (black solid line), Ba  (magenta solid line), and Ra (blue solid lines) targets. \label{fig_SM_S1}}
\end{figure}   
From the DCS, the TCS can be obtained by integrating over the entire solid angle $\Omega$ \cite{wavemech_applications}: 
\begin{equation}
\label{tcs}
    \sigma_{\mbox{\scriptsize total}}=\sum_\ell \sigma_\ell  = \int \frac{d\sigma}{d\Omega} d\Omega=2\pi \int_{0}^{\pi} \frac{d\sigma}{d\Omega} \sin\theta d\theta,
\end{equation}
where $\sigma_\ell$ represent PCS. The elastic cross section analysis is performed using the well-established computational package ELSEPA \cite{salvat2005elsepa}. This package is widely used for calculating scattering parameters for atomic and molecular targets with the electron/positron projectile and has given highly reliable results consistent with experimental data. 

Figure \ref{fig_SM_S1} shows the TCS [Eq. (\ref{tcs})] obtained for Mg, Cd, Ba, and Ra. A cross section has both resonant and non-resonant contributions. The cross section magnitude is found to be the maximum for Ba followed by Ra, Mg and Cd. The resonance is sharper for the former two and relatively weak in the latter two. 

Assuming that the resonant contribution may not affect the diffraction phenomenon, the $V_\text{Eff}$, given by Eq. (\ref{V_avg}), is also computed using the non-resonant total cross-section ($\sigma$) as being the weight factor. As the resonances are sharp, the corresponding resonant partial waves were subtracted from $\sigma$ to avoid their effect without any appreciable change in the non-resonant strength. For instance, in the case of Cd, the partial wave $\ell=1$ was found to be resonant and therefore, that channel was excluded from $\sigma$ while performing the weighted averaging to obtain $V_\text{Eff}$. Thus, the $V_\text{Eff}$ in Eq. (\ref{V_avg}) is computed using $\sigma$ both with and without the effect of resonance.

Finally, Figure \ref{fig_SM_S2} shows a comparison of the TCS of the \textit{e}-Cd elastic scattering, obtained in our experiment and using our Dirac partial wave formalism. The agreement between the theoretical and experimental results is excellent.

\section{Fourier transformation method}

The oscillations (fringes) observed in the DCS are processed using a Fourier Transform (FT) technique. The DCS is first plotted in real ($q$) space and then analyzed in the reciprocal (radial) coordinate space using the Fast Fourier Transform (FFT) algorithm~\cite{press2002art}. To isolate the diffraction signal, the non-oscillatory (non-diffractive) background component is removed by dividing the DCS by a fitted background function. A fitting function of the form $q^x$, where $x$ is carefully adjusted to match the background, is employed for DCS corresponding to each energy. For the FFT calculation, the DCS data are sampled at equal \textit{q}-intervals and adjusted to contain $2^n$ data points by applying zero-padding beyond the considered \textit{q}-range~\cite{press2002art}. The resulting nearly-pure oscillation signal is then fed into the FFT algorithm, with an appropriate window function applied to ensure smoother transformed curves. From the FFT output, the location of the first prominent peak is considered to be the diffractor size. Corresponding to each energy, the diffractor size is then plotted against the incident energy in Fig.\,4 of the main paper.

\renewcommand{\thefigure}{S2}
\begin{figure} [t] 
\includegraphics[width=7.5cm]{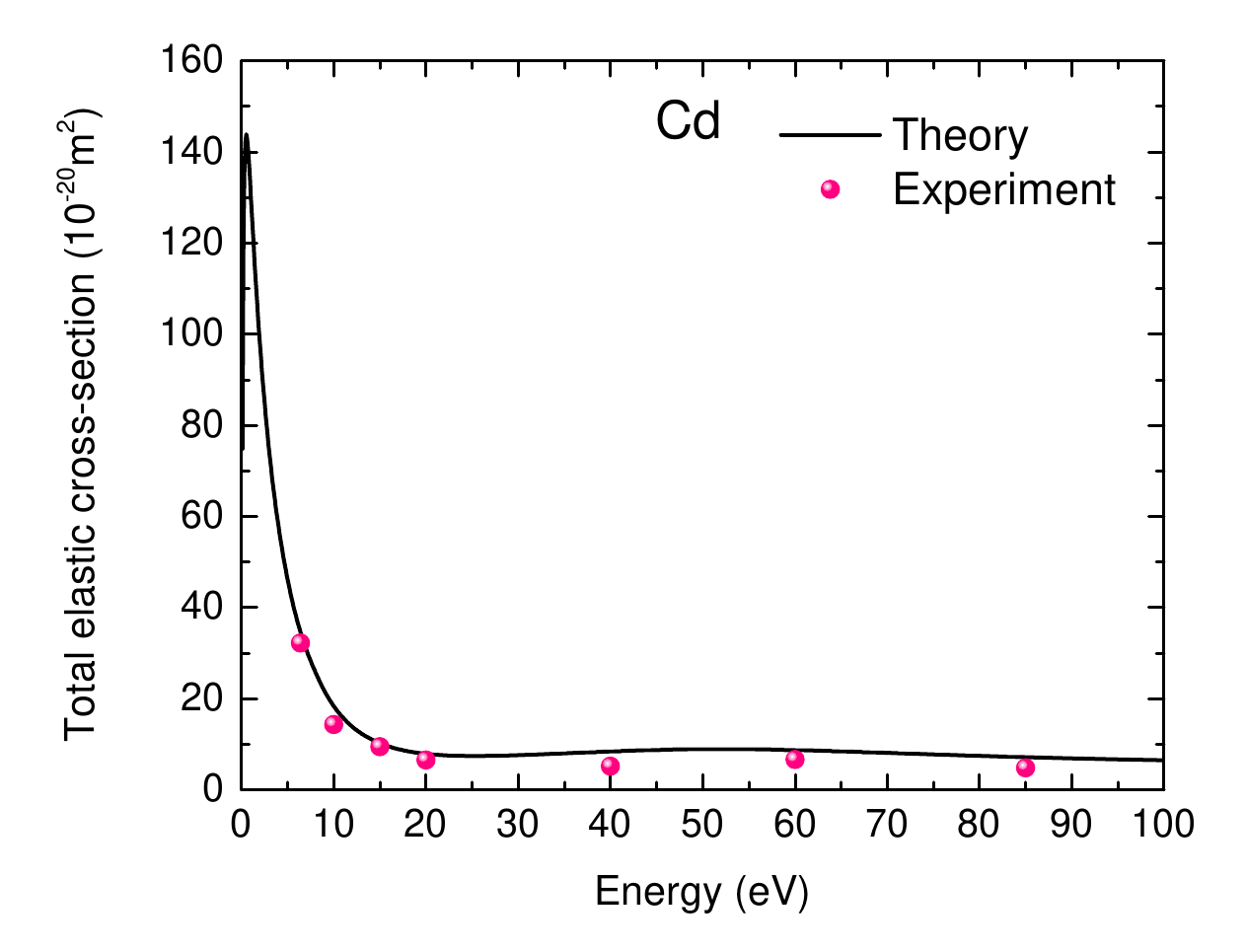}
\caption{TCS of \textit{e}–Cd scattering: comparison between theoretical calculations and experimental data across 0.1 to 100 eV energy range.  \label{fig_SM_S2}}
\end{figure}
\begin{table*}[htbp]
    \renewcommand\thetable{S1}
    \renewcommand{\arraystretch}{1}
    \begin{center}
    \begin{tabular}{|W{c}{1.5cm}|W{c}{1.5cm}|W{c}{1.5cm}|W{c}{1.5cm}|W{c}{1.5cm}|W{c}{1.5cm}|W{c}{1.5cm}|W{c}{1.5cm}|W{c}{1.5cm}|}
     \hline
                 &  \multicolumn{8}{c|}{\textbf{ Elastic differential cross section ($10^{-20}~\text{m}^2$)}}	\\[1ex]
    \cline{2-9}
        Angle    &\textit{E=}3.4 eV &\textit{E=}6.4 eV &\textit{E=}10 eV  &\textit{E=}15 eV  &\textit{E=}20 eV  &\textit{E=}40 eV  &\textit{E=}60 eV  &\textit{E=}85 eV         \\ [1ex]
        (Degree) &(35\%) & (30\%)& (30\%)& (30\%)&(30\%) & (35\%)& (35\%)& (35\%)       \\
    \hline
         6       &       &       &       &       &       &23.000 &       & 27.600        \\[1ex]
         8       &       &       &       &       &       &15.600 &       &               \\[1ex]
         9       &       &       &       &       &       &       &15.600 &                \\[1ex]
        10       &       &36.200 &29.000 &16.600 &       &11.500 &13.400 & 9.510         \\[1ex]
        11       &       &       &       &       &13.700 &       &       &               \\[1ex]
        15       &       &28.400 &18.500 &14.900 &7.820  &4.660  &5.020  & 3.550         \\ [1ex]      
        20       &31.600 &21.300 &11.600 &8.940  &4.000  &1.860  &2.470  & 1.940         \\[1ex]
        30       &22.400 &10.700 &3.950  &1.830  &0.626  &0.284  &0.998  & 1.040         \\ [1ex] 
        35       &       &       &       &       &       &0.156  &       &                \\[1ex]    
        40       &15.100 &5.440  &1.140  &0.216  &0.029  &0.143  &0.450  & 0.438         \\[1ex]    
        42       &       &       &       &       &0.017  &       &       &               \\[1ex]    
        45       &       &       &       &       &       &0.167  &       &               \\[1ex]    
        46       &       &       &       &       &0.025  &       &       &                \\[1ex]    
        50       &9.360  &2.740  &0.265  &0.013  &0.056  &0.190  &0.179  & 0.085         \\[1ex]    
        54       &       &       &       &0.016  &       &       &       &                \\[1ex]
        55       &       &       &       &       &0.089  &0.199  &       &               \\[1ex]
        60       &5.620  &1.350  &0.052  &0.038  &0.123  &0.185  &0.082  & 0.021         \\[1ex]
        70       &3.190  &0.613  &0.017  &0.043  &0.106  &0.111  &0.055  & 0.086         \\[1ex]
        80       &1.690  &0.231  &0.006  &0.024  &0.047  &0.020  &0.044  & 0.125         \\[1ex]
        90       &0.972  &0.081  &0.004  &0.005  &0.008  &0.009  &0.048  & 0.094         \\[1ex]
        92       &       &       &       &       &0.006  &       &       &                \\[1ex]
        96       &       &       &       &       &0.004  &       &       &                \\[1ex]
        100      &0.694  &0.058  &0.004  &0.002  &0.006  &0.070  &0.108  & 0.055         \\[1ex]
        102      &       &       &       &       &0.007  &       &       &               \\[1ex]
        110      &0.567  &0.114  &0.011  &0.002  &0.018  &0.154  &0.182  & 0.054         \\[1ex]
        120      &0.535  &0.221  &0.032  &0.004  &0.013  &0.168  &0.212  & 0.050         \\[1ex]
        126      &       &       &       &       &0.006  &       &       &               \\[1ex]
        130      &0.612  &0.383  &0.067  &0.022  &0.007  &0.121  &0.146  & 0.024         \\[1ex]  
        136      &       &       &       &       &       &       &0.055  &               \\[1ex]
        140      &0.775  &       &0.122  &0.099  &0.043  &0.032  &0.021  & 0.003         \\[1ex]
        142      &       &       &       &       &0.063  &0.022  &       &              \\[1ex]
        144      &       &       &       &       &       &0.017  &       &              \\[1ex]    
        146      &0.846  &       &       &       &       &0.021  &       &  0.003        \\[1ex]   
        148      &       &       &0.201  &       &       &0.029  &0.042  &  0.054        \\[1ex]
        150      &       &       &0.214  &0.267  &0.179  &0.045  &       &  0.078        \\
        \hline
    \end{tabular}
    \end{center}
    \caption{The DCS of Cd, expressed in units of $10^{-20}~\text{m}^2$, are reported for incident electron energies 3.4 eV, 6.4 eV, 10 eV, 15 eV, 20 eV, 40 eV, 60 eV, and 85 eV. The experimental uncertainties associated with each energy value are provided in parentheses. }
    \label{table_S1}
\end{table*}

\section{Numerical data from experiment}

The DCS of \textit{e}-Cd elastic scattering also shows excellent agreement between theory and experiment (Refer to Fig. 1 in the main text). Table \ref{table_S1} presents the numerical data of the DCS from the current experiment. 
\clearpage 

\bibliography{reference-sm}